# A universal model for mobility and migration patterns

Filippo Simini[1,2,3], Marta C. González[4], Amos Maritan[2], and Albert-László Barabási[1,5,6]

[1] *Center for Complex Network Research and Department of Physics, Biology and Computer Science, Northeastern University, Boston, Massachusetts 02115, USA*

[2] *Dipartimento di Fisica "G. Galilei", Università di Padova, CNISM and INFN, via Marzolo 8, 35131 Padova, Italy*

[3] *Institute of Physics, Budapest University of Technology and Economics, Budafoki út 8, Budapest, H-1111, Hungary*

[4] *MIT - Department of Civil and Environmental Engineering, 77 Massachusetts Avenue, Cambridge, Massachusetts 02139, USA.*

[5] *Center for Cancer Systems Biology, Dana-Farber Cancer Institute, Boston, Massachusetts 02115, USA*

[6] *Department of Medicine, Brigham and Women's Hospital, Harvard Medical School, Boston, Massachusetts 02115, USA*

**Introduced in its contemporary form by George Kingsley Zipf[1] in 1946, but with roots that go back to the work of Gaspard Monge[2] in the 18th century, the gravity law[1,3,4] is the prevailing framework to predict population movement[3,5,6], cargo shipping volume[7], inter-city phone calls[8,9], as well as bilateral trade flows between nations[10]. Despite its widespread use, it relies on adjustable parameters that vary from region to region and suffers from known analytic inconsistencies. Here we introduce a stochastic process capturing local mobility decisions that helps us analytically derive commuting and mobility fluxes that require as input only information on the population distribution. The resulting radiation model predicts mobility patterns in good agreement with mobility and transport patterns observed in a wide range of phenomena, from long-term migration patterns to communication volume between different regions. Given its parameter-free nature, the model can be applied in areas where we lack previous mobility measurements, significantly improving the predictive accuracy of most of phenomena affected by mobility and transport processes[11-23].**

In analogy with Newton's law of gravity, the gravity law assumes that the number of individuals $T_{ij}$ that move between locations $i$ and $j$ per unit time is proportional to some power of the population of the source ($m_i$) and destination ($n_j$) locations, and decays with the distance $r_{ij}$ between them as

$$T_{ij} = \frac{m_i^\alpha n_j^\beta}{f(r_{ij})}, \qquad (1)$$

where $\alpha$ and $\beta$ are adjustable exponents and the deterrence function $f(r_{ij})$ is chosen to fit the empirical data. Occasionally $T_{ij}$ is interpreted as the probability rate of individuals of traveling from *i* to *j*, or an effective coupling between the two locations[24]. Despite its widespread use, the gravity law has notable limitations:

i)   We lack a rigorous derivation of (1). While entropy maximization[25] leads to (1) with $\alpha = \beta = 1$, it fails to offer the functional form of $f(r)$.

ii)   Lacking theoretical guidance, practitioners use a range of deterrence functions (power law or exponential) and up to nine parameters to fit the empirical data[5,7,8,11,14].

iii)   As (1) requires previous traffic data to fit the parameters [$\alpha, \beta, ...$], it is unable to predict mobility in regions where we lack systematic traffic data, areas of major interest in modeling of infectious diseases.

iv)   The gravity law has systematic predictive discrepancies. Indeed, in **Fig. 1a** we highlight two pairs of counties with similar origin and destination populations and comparable distance, so according to (1) the flux between them should be the same. Yet, the US census (see SI) documents an order of magnitude difference between the two fluxes: only 6 individuals commute between the two Alabama counties, while 44 in Utah.

v)   Equation (1) predicts that the number of commuters increases without limit as we increase the destination population $n_j$, yet the number of commuters cannot exceed the source population $m_i$, highlighting the gravity law's analytical inconsistency (see SI, Sect. 4).

vi)   Being deterministic, the gravity law cannot account for fluctuations in the number of travelers between two locations.

Motivated by these known limitations, alternative approaches like the intervening opportunity model[26] or the random utility model[27] (SI, Sect. 7) have been proposed.

While derived from first principles, these models continue to contain context specific tunable parameters, and their predictive power is at best comparable to the gravity law[28].

Here we introduce a modelling framework that relies on first principles and overcomes the problems (i) – (vi) of the gravity law. While commuting is a daily process, its source and destination is determined by job selection, a decision made over longer timescales. Using the natural partition of a country into counties (for which commuting data are collected), we assume that job selection consists of two steps (**Fig. 1 b, c**):

1) An individual seeks job offers from all counties, including his/her home county. The number of employment opportunities in each county is proportional to the resident population, $n$, assuming that there is one job opening for every $n_{jobs}$ individuals. We capture the benefits of a potential employment opportunity with a single number, $z$, randomly chosen from distribution $p(z)$ where $z$ represents a combination of income, working hours, conditions, etc. Thus, each county with population $n$ is assigned $n/n_{jobs}$ random numbers, $z_1, z_2, \ldots, z_{[n/n_{jobs}]}$, accounting for the fact that larger a county's population, the more employment opportunities it offers.

2) The individual chooses the *closest* job to his/her home, whose benefits $z$ are *higher* than the best offer available in his/her home county. Thus lack of commuting has priority over the benefits, i.e. individuals are willing to accept lesser jobs closer to their home.

This process, applied in proportion to the resident population in each county, assigns work locations to each potential commuter, which in turn determines the daily commuting fluxes across the country. The model has three unknown parameters: the benefit distribution $p(z)$, the job density $n_{jobs}$, and the total number of commuters, $N_c$. We show, however, that the commuting fluxes $T_{ij}$ are independent of $p(z)$ and $n_{jobs}$, and the remaining free parameter, $N_c$, does not affect the flux distribution, making the model parameter free. As the model can be formulated in terms of radiation and absorption processes (see SI, Sec. 2), we will refer to it as the *radiation model*. To analytically predict the commuting fluxes we consider locations $i$ and $j$ with population $m_i$ and $n_j$ respectively, at distance $r_{ij}$ from each other, and we denote with $s_{ij}$ the total population in the circle of radius $r_{ij}$ centered at $i$ (excluding the source and destination population). The average flux $T_{ij}$ from $i$ to $j$, as predicted by the radiation model (see SI, sect. 2), is

$$\langle T_{ij} \rangle = T_i \frac{m_i n_j}{(m_i + s_{ij})(m_i + n_j + s_{ij})}, \qquad (2)$$

which is independent of both $p(z)$ and $n_{jobs}$. Hence (2) represents the fundamental equation of the radiation model, the proposed alternative to the gravity law (1). Here $T_i \equiv \sum_{j \neq i} T_{ij}$ is the total number of commuters that start their journey from location $i$, which is proportional to the population of the source location, hence $T_i = m_i(N_c/N)$, where $N_c$ is the total number of commuters and $N$ is the total population in the country (**Fig. 2g**).

Equation (2) resolves the limitations (i) – (vi) of the gravity law: it has a rigorous derivation (resolving (i)) and has no free parameters (bypassing (ii) and (iii)). To understand the origin of (iv), we note that a key difference between the radiation model (2) and the gravity law (1) is that the variable of (2) is not the distance $r_{ij}$, but $s_{ij}$. Thus the commuting flux depends not only on $m_i$ and $n_j$ but also on the population $s_{ij}$ of the region surrounding the source location. For uniform population density $s_{ij} \sim m_i r_{ij}^2$ and $n = m$, (2) reduces to the gravity law (1) with $f(r) = r^\gamma$, $\gamma = 4$ and $\alpha + \beta = 1$. The non-uniform population density, however, is key to resolving problem (iv): Eq. (2) predicts an order of magnitude difference in Alabama and Utah, in line with the census data (see **Fig. 1a**). Indeed the population density around Utah is significantly lower than the United States average, thus work opportunities within the same radius are ten times smaller in Utah than in Alabama, implying that commuters in Utah have to travel farther to find comparable employment opportunities. Note also that Eq. (2) predicts that the number of travelers leaving from a location with population $m$ to one with $n \to \infty$ saturates at $T_{n \to \infty} = \frac{m^2}{(m+s)} + O\left(\frac{1}{n}\right) \leq m$, resolving the unphysical divergence highlighted in (v). Finally, $T_{ij}$ in the radiation model is a stochastic variable, predicting not only the average flux between two locations (2), but also its variance (see SI Sect. 2), resolving the problem (vi).

To explore the radiation model's ability to predict the correct commuting patterns, in **Figure 2a** we show the commuting fluxes with more than ten travelers originating from New York County. The destinations predicted by the gravity law[14] are all within 400 km from the origin, missing all long distance and many medium distance trips. The gravity law's local performance is equally poor: within the State of New York it grossly overestimates fluxes in the vicinity of New York City and underestimates the fluxes in the rest of the state (**Fig. 2a, right column**). The radiation model offers a more realistic approximation to the observed commuting patterns, both nationally and statewide (**Fig. 2a, bottom panels**). To quantify the observed differences, we compare the measured and the predicted non-zero commuting fluxes for all pairs of counties in the United States. We find that both standard implementations of the gravity law[11,14] ($f(r) = r^\gamma$ and $f(r) = e^{dr}$) significantly underestimate the high flux commuting patterns, often by an order of magnitude or more (**Fig. 2b, c**). In contrast, the average fluxes predicted by the radiation model are within the error bars despite the observed six orders of magnitude span in commuting fluxes (**Fig. 2d**).

The systematic failure of the gravity law is particularly evident if we measure the probability $P_{\text{dist}}(r)$ of a trip between locations at distance $r$ (**Fig. 2e**), and the probability of trips towards a destination with population $n$, $P_{\text{dest}}(n)$ (**Fig. 2f**). For $P_{\text{dist}}(r)$ the radiation model clearly follows the peak around 40 km in the census data. The prediction based on the gravity law lacks this peak and thus it overestimates by three orders of magnitude the number of short distance trips. Similarly, the gravity law overestimates the low $n$ values of $P_{\text{dest}}(n)$ by nearly an order of magnitude.

Another important mobility measure is the conditional probability $p(T|m,n,r)$ to observe a flux of $T$ individuals from a location with population $m$ to a location with population $n$ at a distance $r$. The gravity law predicts a highly peaked $p(T|m,n,r)$ distribution around the average $\langle T \rangle_{mnr} = \sum_T p(T|m,n,r)T$ (**Fig. 2 h, i, j**), because, according to (1) pairs of locations with the same ($m$, $n$, $r$) have the same flux. In contrast the radiation model predicts a broad $p(T|m,n,r)$ distribution, in reasonable agreement with the data.

To show the generality of the model in **Fig. 3** we test its performance for four socio/economic phenomena: hourly travel patterns, migrations, communication patterns, and commodity flows. We find that the radiation model offers an accurate quantitative description of mobility and transport spanning a wide range of time scales (hourly mobility, daily commuting, yearly migrations), capturing diverse processes (commuting, intra-day mobility, call patterns, trade), collected via a wide range of tools (census, mobile phones, tax documents) on different continents (America, Europe). The agreement with data of such diverse nature is somewhat surprising, suggesting that the hypotheses behind the model capture fundamental decision mechanisms that, directly or indirectly, are relevant to a wide span of mobility and transport-driven processes.

To illustrate the effect of the heterogeneous population distribution on commuting fluxes, in **Fig S8 a-f** we show the commuting landscape generated by (2) from the perspective of two individuals, one in Davis county, UT, and the other in Clayton County, GA, with comparable populations 238,994 and 236,517, respectively. If the population was uniformly distributed, the landscape seen by a potential employee would be simple: the farther is a job, the less desirable it is (**Fig. S8 a,d**). Yet, the observed variations in population density significantly alter the local commuting landscape, as shown in **Fig. S8 b,e** where we colored the US counties based on their distance to the commuter's home county (**Fig. S8 a,d**) and then moved them closer or further from the origin so that the new distance reflects the true likelihood of representing a commuting destination.

Despite the observed differences in the perspective of individual commuters, the radiation model helps us uncover a previously unsuspected scale-invariance in commuting patterns. Indeed, according to (2), the probability of one trip from *i* to *j* (equal to $T_{ij}/T_i$) is scale invariant under the transformation $m_i \to \lambda m_i$, $n_j \to \lambda n_j$, and $s_{ij} \to \lambda s_{ij}$. Empirical evidence for this statistical self-similarity is offered in **Fig. 4a,b** (see also SI sect. 8).

In summary, the superior performance of the radiation model can significantly improve the accuracy of predictive tools in all areas affected by mobility and transport processes[11,12], from epidemiology[13] and spreading processes[17], to urban geography[18-21], and flow of resources in economics[22]. The parameter-free modeling platform we introduced can predict commuting and transport patterns even in areas where such data is not collected systematically, as it relies only on population densities, which is relatively accurately estimated throughout the globe.

Despite its superior performance the radiation model can absorb further improvements. For example, consider the fact that an individual has a home-field advantage when searching for jobs in the home county, being more familiar with local employment opportunities. We can incorporate this by adding $\varepsilon/n_{jobs}$ additional employment opportunities to his/her home county, achieved through an effective increase $m \to m + \varepsilon$ of the home county population, so the adjusted law is now invariant under the $(m+\varepsilon) \to \lambda(m+\varepsilon)$ and $s \to \lambda s$ transformation. We find that the rescaling of the commuting probability improves dramatically (**Fig. 4c**), indicating that the home field advantage offers an effective boost in employment opportunities that is equivalent with an additional $\varepsilon = 35,000$ individuals in the home county population. Furthermore, the adjusted radiation model displays a better or equally good agreement with the real data in all tested measures (**Fig S6**), demonstrating that Eq. (2) is not a rigid end point of our approach, but offers a platform that can be improved upon in specific environments.

**Acknowledgments** We thank J. P. Bagrow, A. Fava, F. Giannotti, Yu-Ru Lin, J. Menche, Z. Néda, D. Pedreschi, D. Wang, G. Wilkerson, and D. Bauer, and for many useful discussions, and especially Neil Ferguson for prompting us to look carefully into the gravity law. AM and FS acknowledge Cariparo foundation for financial support. This work was supported by the Network Science Collaborative Technology Alliance sponsored by the US Army Research Laboratory under Agreement Number W911NF-09-2-0053; the Office of Naval Research under Agreement Number N000141010968; the Defense Threat Reduction Agency awards WMD BRBAA07-J-2-0035 and BRBAA08-Per4-C-2-0033; and the James S. McDonnell Foundation 21st Century Initiative in Studying Complex Systems.

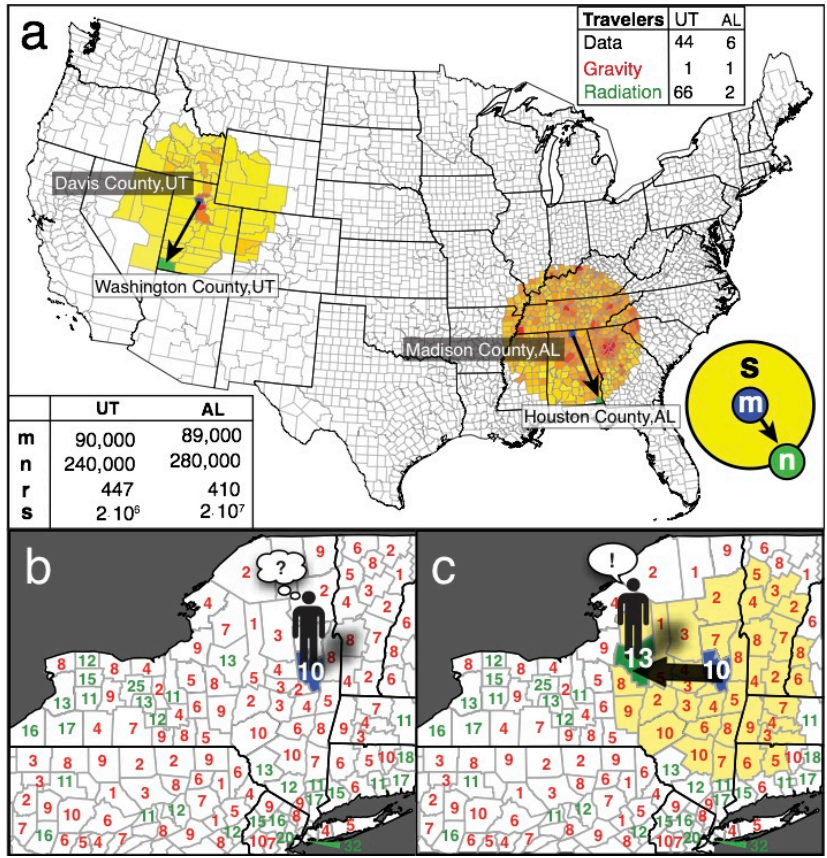

**Figure 1**

**Figure 1**

**The radiation model.** (*a*) To demonstrate the limitations of the gravity law we highlight two pairs of counties, one in Utah (UT) and the other in Alabama (AL), with similar origin (*m*, blue) and destination (*n*, green) populations and comparable distance *r* between them (see bottom left table). The gravity law predictions were obtained by fitting Eq. (1) to the full commuting dataset, recovering the parameters [$\alpha$, $\beta$, $\gamma$] = [0.30, 0.64, 3.05] for $r < 119$ km, and [0.24, 0.14, 0.29] for $r > 119$ km of Ref. 14. The fluxes predicted by (1) are the same because the two county pairs have similar *m*, *n*, and *r* (top right table). Yet the US census 2000 reports a flux that is an order of magnitude greater between the Utah counties, a difference correctly captured by the radiation model (*b, c*). The definition of the radiation model: (*b*) An individual (e.g. living in Saratoga County, NY) applies for jobs in all counties and collects potential employment offers. The number of job opportunities in each county (*j*) is $n_j/n_{jobs}$, chosen to be proportional to the resident population $n_j$. Each offer's attractiveness (benefit) is represented by a random variable with distribution *p(z)*, the numbers placed in each county representing the best offer among the $n_j/n_{jobs}$ trials in that area. Each county is marked in green (red) if its best offer is better (lower) than the best offer in the home county (here $z = 10$). (*c*) An individual accepts the closest job that offers better benefits than his home county. In the shown configuration the individual will commute to Oneida County, NY, the closest county whose benefit $z = 13$ exceeds the home county benefit $z = 10$. This process is repeated for each potential commuter, choosing new benefit variables *z* in each case.

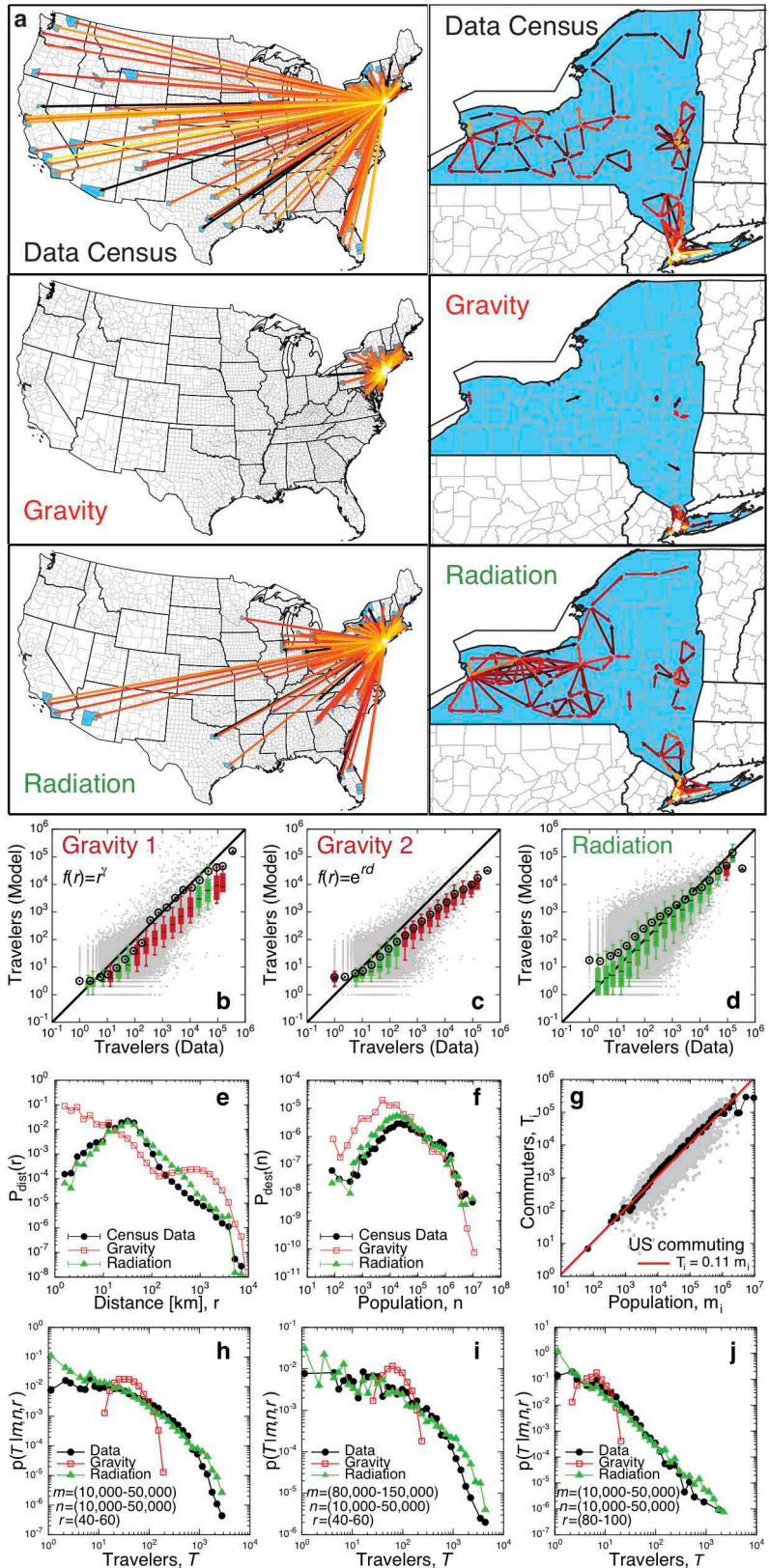

**Figure 2**

**Figure 2**

**Comparing the predictions of the radiation model and the gravity law.** (*a*) National mobility fluxes with more than ten travelers originating from New York County (left panels) and the high intensity fluxes (over 1,100 travelers) within the State of New York (right panels). Arrows represent commuters fluxes, the color capturing flux intensity: black = 10 individuals (fluxes below ten travelers are not shown for clarity), white > 10,000 individuals. The top panels display the fluxes reported in US census 2000, the central panels display the fluxes fitted by the gravity law with[14] $f(r) = r^\gamma$, and the bottom panels display the fluxes predicted by the radiation model. (*b, c, d*): Comparing the measured flux, $T_{ij}^{data}$, with the predicted flux, $T_{ij}^{GM}$ and $T_{ij}^{Rad}$, for each pair of counties. We compare the census data with two formulations of the gravity law, (*b*) $f(r) = r^\gamma$ and (*c*) $f(r) = e^{dr}$, and (*d*) with the radiation model. Gray points are scatter plot for each pair of counties. A box is colored green if the line Y=X lies between the 9th and the 91st percentiles in that bin, it is red otherwise. The black circles correspond to the mean number of predicted travelers in that bin. (*e*) Probability of a trip between two counties that are at distance *r* kms from each other, $P_{dist}(r)$. (*f*) Probability of a trip towards a county with population *n*, $P_{dest}(n)$. (*g*), The number of commuters in a county, $T_i$, is proportional to its population, $m_i$. (*h, i, j*) Conditional probability p(*T*|*m,n,r*) to observe a flow of *T* individuals from a location with population *m* to a location with population *n* at a distance *r* for three triplets (*m,n,r*). The gravity law predicts a highly peaked distribution around the average value $\langle T \rangle_{mnr}$, in disagreement with census data and the radiation model, which both display a broad distribution.

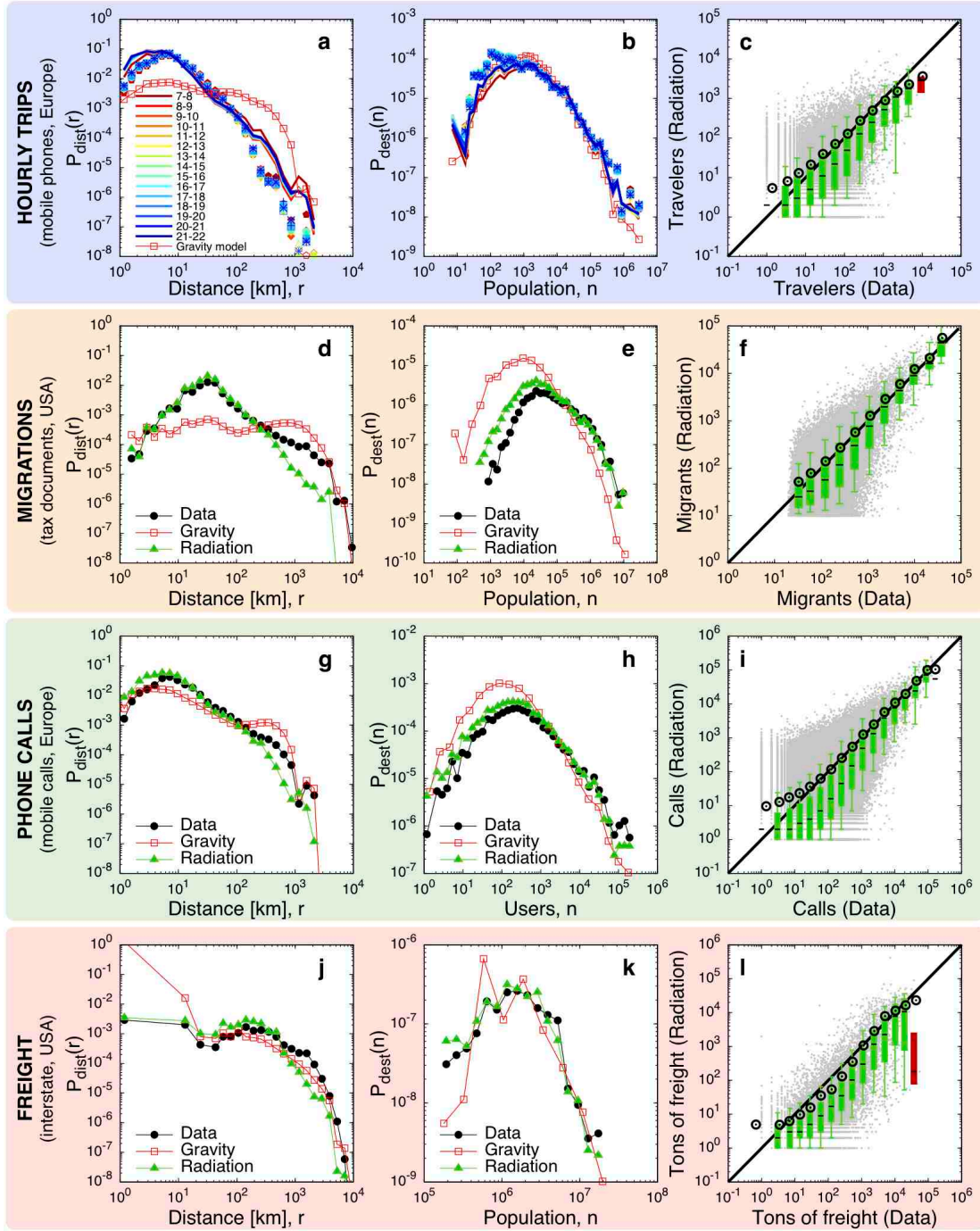

**Figure 3**

**Beyond commuting.** (*a-c*): Testing the radiation model on hourly trips extracted from a mobile phone database of a western European country. The anonymized billing records[29,30] covers the activity of approximately 10M subscribers. We analyzed a six-month period, recording the user locations with tower resolution hourly between 7am and 10pm, identifying all trips between municipalities. (*a*) Probability of a trip between two municipalities at distance *r*, $P_{dist}(r)$, shown for 14 hourly time intervals. Radiation model predictions are solid lines; gravity law's aggregated fit over 24 hours is a red line with empty squares. (*b*) Probability of a trip toward a municipality with population *n*, $P_{dest}(n)$. (*c*) Comparing the measured flux $T_{ij}^{data}$, with the predicted flux, $T_{ij}^{Rad}$, for each pair of municipalities with $T_{ij}^{data}, T_{ij}^{Rad} > 0$, for commuting trips extracted by identifying each user's home and workplace from the locations where the user made the most calls. (*d, e, f*): Testing (2) on long-term migration patterns, capturing the number of individuals that relocated from one US county to another during tax years 2007-2008 as reported by the US Internal Revenue Service. (*g, h, i*) Phone call volume between municipalities extracted from the anonymized mobile phone database. The number of phone calls between users living in different municipalities during a period of four weeks resulted in 38,649,153 calls placed by 4,336,217 users. We aggregated the data to obtain the total number of calls between every pair of municipalities. (*j, k, l*)**:** Commodity flows in the US extracted from the Freight Analysis Framework (FAF), which offers a comprehensive picture of freight movement among US states and major metropolitan areas by all modes of transportation. For each dataset we measured the quantities discussed in *a-c*.

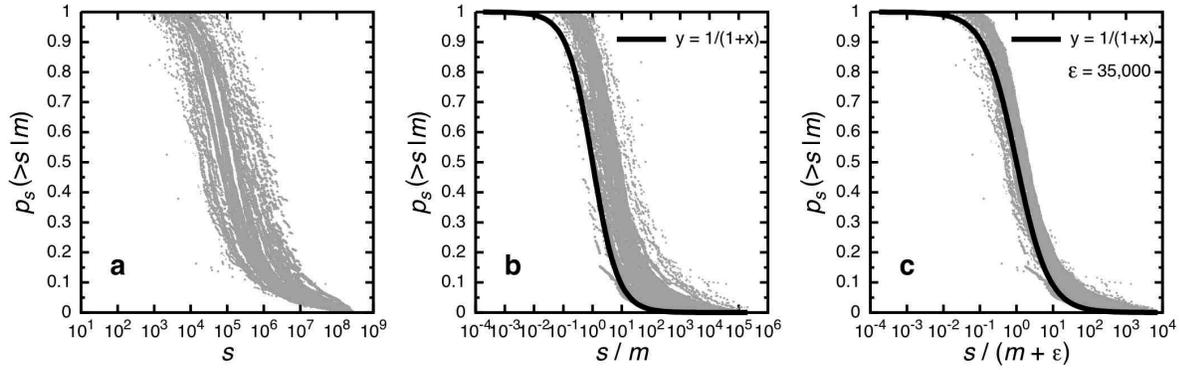

**Figure 4**

**Unveiling the hidden self-similarity in human mobility**. (*a*), The probability $p_s(\geq s\,|\,m)$ to observe a trip from a location with population *m* to a destination in the region beyond a population *s* from the origin (*m* varies between 200 and 2,000,000). (*b*), According to the radiation model $p_s(\geq s\,|\,m) = 1/(1 + s/m)$, a homogeneous function of the ratio $s/m$. Plotting $p_s(\geq s\,|\,m)$ vs $s/m$ the curves approach the theoretical result $y = 1/(1+x)$. (*c*), The collapse improves if we account for the home field advantage in job search by always adding $\varepsilon = 35,000$ to the population of the commuter's home county.

# Supplementary Information

# A universal model for mobility and migration patterns

Filippo Simini, Marta C. González, Amos Maritan, and Albert-László Barabási

**Contents:**

# 1 Datasets

## 1.1 US commuting

Data on commuting trips between United States counties are available online at http://www.census.gov/population/www/cen2000/commuting/index.html.

The files were compiled from Census 2000 responses to the long-form (sample) questions on where individuals worked. The files provide data at the county level for residents of the 50 states and the District of Columbia (DC). The data contain information on 34,116,820 commuters in 3,141 counties.

## 1.2 US migrations

United States population migration data for 2007-2008 are available online at http://www.irs.gov/taxstats/article/0,,id=212695,00.html.

The main source of area-to-area migration data in the United States is the Statistics of Income Division (SOI) of the Internal Revenue Service (IRS), which maintains records of all individual income tax forms filed in each year. The Census Bureau is allowed access to tax returns, extracted from the IRS Individual Master File (IMF), which contains administrative data collected for every Form 1040, 1040A, and 1040EZ processed by the IRS. Census determines who in the file has, or has not, moved. To do this, first, coded returns for the current filing year are matched to coded returns filed during the prior year. The mailing addresses on the two returns are then compared to one another. If the two are identical, the return is labeled a "non-migrant." If any of the above information changed during the prior 2 years, the return is considered a mover.

## 1.3 US Commodity flow

FAF[3] data on commodity flow are available online at the following website http://ops.fhwa.dot.gov/freight/freight_analysis/faf/index.htm.

The Freight Analysis Framework (FAF) integrates data from a variety of sources to create a comprehensive picture of freight movement among US states and major metropolitan areas by all modes of transportation. With data from the 2007 Commodity Flow Survey and additional sources, FAF version 3 (FAF$^3$) provides estimates for tonnage and value, by commodity type, mode, origin, and destination for 2007.

The data consists on 5,712,385 total tonnage shipped among 123 states and Major Metropolitan Areas (MMA).

## 1.4 Mobile phone database

We used a set of anonymized billing records from a European mobile phone service provider[29,30,31]. The records cover over 10M subscribers within a single country over 4 years of activity. Each billing record, for voice and text services, contains the unique anonymous identifiers of the caller placing the call and the callee receiving the call; an identifier for the cellular antenna (tower) that handled the call; and the date and time when the call was placed. Coupled with a dataset describing the locations (latitude and longitude) of cellular towers, we have the approximate location of the caller when placing the call. In order to understand whether the radiation model could also describe hourly trips in addition to commuting trips, we analyzed the phone data during a period of six months, recording the locations of users between 7am to 10pm. As the cell phone company's market share is not uniform across the country, in order to have the necessary demographic information we use municipalities and not cell towers as locations. We define a trip from municipality $i$ to municipality $j$ if we recorded the same user in municipality $i$ at time $t$ and in municipality $j$ at time $t+1$, where $t$ is a daytime one-hour time interval (i.e. $t = 7, 8, …, 21, 22$). We aggregated all trips from hour $t$ to hour $t+1$ observed during the six-month period, generating the hourly commuting flows among municipalities.

**1.5 Call patterns**

Using the mobile phone database described above, we also extracted the number of phone calls between users living in different municipalities during the same period, resulting in a total of 38,649,153 calls placed by 4,336,217 users.

**2 The radiation model: analytical results**

Here we propose a more general description of the radiation model, which is completely equivalent to the formulation given in the main text. We use an analogy with radiation emission and absorption processes, widely studied in physical sciences[32]. Imagine the location of origin, $i$, as a source emitting an outgoing flux of identical and independent units (particles). We define the emission/absorption process through the following two steps:

1) We associate to every particle, $X$, emitted from location $i$ a number, $z_X^{(i)}$, that represents the *absorption threshold* for that particle. A particle with large threshold is less likely to be absorbed. We define $z_X^{(i)}$ as the maximum number obtained after $m_i$ random extractions from a preselected distribution, $p(z)$ ($m_i$ is the population in location $i$). Thus, on average, particles emitted from a highly populated location have a higher absorption threshold than those emitted from a scarcely populated location. We will show below that the particular choice of $p(z)$ do not affect the final results.

2) The surrounding locations have a certain probability to absorb particle $X$: $z_X^{(j)}$ represents the *absorbance* of location $j$ for particle $X$, and it is defined as the maximum of $n_j$ extractions from $p(z)$ (remember that $n_j$ is the population in location $j$). *The particle is absorbed by the closest location whose absorbance is greater than its absorption threshold*.

By repeating this process for all emitted particles we obtain the fluxes across the entire country.

We can calculate the probability of one emission/absorption event between any two locations, and thus obtain an analytical prediction for the flux between them. Let $P(1|m_i, n_j, s_{ij})$ be the probability that a particle emitted from location $i$ with population $m_i$ is absorbed in location $j$ with population $n_j$, given that $s_{ij}$ is the total population in all locations (except $i$ and $j$) within a circle of radius $r_{ij}$ centered at $i$ ($r_{ij}$ is the distance between $i$ and $j$). According to the radiation model, we have

$$P(1|m_i, n_j, s_{ij}) = \int_0^\infty dz \; P_{m_i}(z) P_{s_{ij}}(<z) P_{n_j}(>z) \tag{S1}$$

where $P_{m_i}(z)$ is the probability that the maximum value extracted from $p(z)$ after $m_i$ trials is equal to $z$:

$$P_{m_i}(z) = \frac{dP_{m_i}(<z)}{dz} = m_i p(<z)^{m_i - 1} \frac{dp(<z)}{dz}.$$

Similarly, $P_{s_{ij}}(<z) = p(<z)^{s_{ij}}$ is the probability that $s_{ij}$ numbers extracted from the $p(z)$ distribution are *all* less than $z$; and $P_{n_j}(>z) = 1 - p(<z)^{n_j}$ is the probability that among $n_j$ numbers extracted from $p(z)$ *at least one* is greater than $z$.

Thus Eq. (S1) represents the probability that one particle emitted from a location with population $m_i$ is not absorbed by the closest locations with total population $s_{ij}$, and is absorbed in the next location with population $n_j$. After evaluating the above integral, we obtain

$$P(1|m_i,n_j,s_{ij}) = m_i \int_0^\infty dz \frac{dp(<z)}{dz}\left[p(<z)^{m_i+s_{ij}-1} - p(<z)^{m_i+n_j+s_{ij}-1}\right] = m_i\left[\frac{1}{m_i+s_{ij}} - \frac{1}{m_i+n_j+s_{ij}}\right]$$

$$P(1|m_i,n_j,s_{ij}) = \frac{m_i n_j}{(m_i+s_{ij})(m_i+n_j+s_{ij})} \quad (S2)$$

Eq. (S2) is independent of the distribution $p(z)$ and is invariant under rescaling of the population by the same multiplicative factor ($n_{\text{jobs}}$).

The probability $P(T_{i1},T_{i2},...,T_{iL})$ for a particular sequence of absorptions, $(T_{i1},T_{i2},...,T_{iL})$, of the particles emitted at location $i$ is given by the multinomial distribution:

$$P(T_{i1},T_{i2},...,T_{iL}) = \prod_{j \neq i} \frac{T_i!}{T_{ij}!} p_{ij}^{T_{ij}} \qquad \text{with} \sum_{j \neq i} T_{ij} = T_i$$

(S3)

where $T_i$ is the total number of particles emitted by location $i$, and $p_{ij} \equiv P(1|m_i,n_j,s_{ij})$.

The distribution (S3) is normalized because $\sum_{j \neq i} p_{ij} = m_i \sum_{j \neq i}\left[\frac{1}{(m_i+s_{ij})} - \frac{1}{(m_i+n_j+s_{ij})}\right] = 1$.

The probability that exactly $T_{ij}$ particles emitted from location $i$ are absorbed in location $j$ is obtained by marginalizing probability (S3):

$$P(T_{ij} | m_i, n_j, s_{ij}) = \sum_{\substack{\{T_{ik}: k \neq i, j; \\ \sum_{k \neq i} T_{ik} = T_i\}}} P_i(T_{i1}, T_{i2}, ..., T_{ij}, ..., T_{iL}) = \frac{T_i!}{T_{ij}!(T_i - T_{ij})!} p_{ij}^{T_{ij}} (1 - p_{ij})^{T_i - T_{ij}} \quad (S4)$$

that is a binomial distribution with average

$$\langle T_{ij} \rangle \equiv T_i p_{ij} = T_i \frac{m_i n_j}{(m_i + s_{ij})(m_i + n_j + s_{ij})} \quad (S5)$$

and variance $T_i p_{ij}(1 - p_{ij})$.

The proposed radiation model might provide further insights on the problem of defining human agglomerations. For example, the US Census Bureau uses the number of commuters between counties as the basis to define Metropolitan Statistical Areas in the USA. Finding a practical way to define the boundaries of a city is important also because it represents a key difficulty in understanding regularities such as Zipf's law or Gibrat's Law of proportionate growth, since different definitions of cities give rise to different results[33,34].

### 3 The case of uniform population distribution

Comparing the radiation model, Eq. (S5), with the gravity law, Eq. (1), the most noticeable difference is the absence of variable *r*, the distance between the locations of origin and destination, and the introduction of a new variable, *s*, the total population within a circle of radius *r* centred in the origin.

In the particular case of a uniform population distribution, however, we can perform a change of variable and write the radiation model, Eq. (S5), as a function of the distance *r*, as in the gravity law, Eq. (1). If the population is uniformly distributed, then $n = m$ and $s(r) = m\pi r^2$ and the average number of travelers is given by:

$$T_{(m,n,s)} = mP(1|m,n,s) = \frac{m^2 n}{(m+s)(m+n+s)} = \frac{m}{(1+s/m)(2+s/m)}$$

With the change of variable we get

$$T_{(m,n,r)} = \frac{m}{(1+\pi r^2)(2+\pi r^2)} \approx \frac{m}{r^4}, \quad (S6)$$

obtaining a gravity law with $f(r) = r^\gamma$, $\gamma = 4$ and $\alpha + \beta = 1$, which is the same form as the one chosen in Ref. 14, although the value of parameters $\alpha, \beta$ and $\gamma$ are different. However, it is important to note that the assumption of uniform population distribution used to derive Eq. (S6) is not fulfilled in reality, as can be seen in **Fig. S1**, where we plot the distribution of exponents $\mu$, obtained by fitting the data with $s(r) \propto r^\mu$ for $r < 119$ km. The mean value (with a large variance) is found to be $\mu = 2.53$ and not 2, as we would expect for a uniform spatial population distribution. Therefore a gravity law with $f(r) = r^\gamma$ cannot be derived from that argument, and thus there is no reason to prefer it to other functions with an equal number of free parameters that can provide a comparable agreement with the data (one commonly used version is for example[11] $f(r) = e^{dr}$). We will show in Section 6 that from the radiation model it is possible to derive the gravity law in Eq. (S6) under weaker assumptions on the population distribution.

The variable $s$, also called *rank*, has been previously recognized as a relevant quantity in different mobility and social phenomena: it has been first introduced in the intervening opportunity model[26] to describe migrations (see SI sect. 7), and recently it has found applications in the study of social-network friendship[35], and in the analysis of urban movements of Foursquare's users in different cities[36].

**4 Asymptotic limits**

In order to test the analytic self-consistency of the gravity and radiation models it is instructive to calculate the asymptotic limits on the number of trips when the populations in the origin or destination increase. Consider for example the case in which the population of the origin location, $m$, becomes very large compared to $n$ and $s$ ($m >> n,s$). The gravity law (1) predicts that the number of trips, $T$, will diverge to infinity. This is not realistic, because the location of destination cannot offer jobs to an unlimited number of commuters. On the contrary, the radiation model, Eq. (S4), predicts that the number of travelers saturates at $T_{m \to \infty} = n + O\left(\frac{1}{m^2}\right)$, and that the other commuters will travel to farther locations.

The inconsistency of the gravity law is even more evident when we consider the limit of large population in the location of destination ($n >> m,s$). Again Eq. (1) predicts that the number of trips, $T$, will increase without limit, which is impossible considering that the population in the origin is unchanged and the number of commuters cannot exceed the total population. The radiation model, instead, predicts that the number of travelers saturates at $T_{n \to \infty} = \frac{m^2}{(m+s)} + O\left(\frac{1}{n}\right) \leq m$.

These considerations provide further support to our thesis that *any* form of the gravity law (1) is inherently inappropriate to satisfactorily describe commuting fluxes.

**5 Determination of the gravity law's parameters**

Here we describe the method that we used to find the best values for the gravity law parameters. The gravity law assumes that the average number of travelers between two locations, $T_{ij}^{GM}$, can be expressed as a function of the two populations and the distance as $T_{ij}^{GM} = C \frac{m_i^\alpha n_j^\beta}{r_{ij}^\gamma}$. Taking the logarithm on both sides we obtain $\ln(T_{ij}^{GM}) = \ln(C) + \alpha \ln(m_i) + \beta \ln(n_j) - \gamma \ln(r_{ij})$. It is then customary in many studies to use $\ln T_{ij}^{data}$, the logarithm of the number of travelers in the data, to estimate the values of $\ln C, \alpha, \beta, \gamma$ through a least-squares regression analysis[1].

In order to obtain more accurate results, following Ref. 14 we use a nine-parameter form of the gravity law, in which short and long trips are fitted separately. The parameters are two sets of $[C, \alpha, \beta, \gamma]$, one for short and one for long trips, plus $\delta$, the cutoff distance that defines short and long trips (i.e. short if $r_{ij} < \delta$, long otherwise).

We define the error function as the mean square deviation of the logarithms of trips:

$$E = \frac{1}{N} \sum_{\{i,j: i \neq j\}} [\ln(T_{ij}^{GM}) - \ln(T_{ij}^{data})]^2, \tag{S7}$$

where the sum extends over all locations pairs $(i, j)$ such that $T_{ij}^{data}, T_{ij}^{GM} > 0$, and $N$ is the total number of county pairs considered. We finally select the particular set of the nine parameters that minimizes $E$ in Eq. (S7).

Moreover, we checked that the power law deterrence function, $f(r_{ij}) = r_{ij}^\gamma$, provides a better fit compared to the exponential deterrence function, $f(r_{ij}) = e^{dr_{ij}}$. Indeed, we found that the best-fit parameters obtained with the power law deterrence function always provide smaller errors $E$.

---

[1] As pointed out in Ref. 37 there are other ways to determine the values of the best-fit parameters. Here we will only consider the most widely used approach for "Newtonian" gravity laws, of the type of Eq. (1), defined above.

The sensitivity of the gravity law to the number of zero fluxes is known to be one of the main factors to determine the quality of the gravity fit. Indeed, for freight transportation, in which the gravity law provides its better prediction, 97.5% of all possible pairs of locations have a non-zero flux, compared to 3.2% of phone calls, 1.65% of commuting fluxes, and 0.8% of migrations.

## 6 Relationship between the gravity law and the radiation model

To understand the relationship between the gravity law and the radiation model, let us assume, in line with our empirical observations, that the radiation model provides a good approximation to the data, i.e. $T_{ij}^{data} \approx T_{ij}^{Rad} = \frac{m_i^2 n_j}{(m_i + s_{ij})(m_i + n_j + s_{ij})}$. To identify the conditions under which the gravity law also offers a good fit to the data, let us minimize the following error function:

$$E = \frac{1}{N} \sum_{\{i,j: i \neq j\}} \left[ \ln\left(C \frac{m_i^\alpha n_j^\beta}{r_{ij}^\gamma}\right) - \ln\left(\frac{m_i^2 n_j}{(m_i + s_{ij})(m_i + n_j + s_{ij})}\right) \right]^2, \qquad (S8)$$

i.e. determine the conditions (and the values of [$\alpha$, $\beta$, $\gamma$]) under which the mean square deviation between gravity's and radiation's predictions is minimized.

We can write the variables $m_i$ and $n_j$ as deviations from the *local* average population, $\overline{m} \equiv (m_i + n_j + s_{ij})/N_{ij}$, where $N_{ij}$ is the number of locations in a circle of radius $r_{ij}$ centered in $i$ ($r_{ij}$ is the distance between $i$ and $j$), and $(m_i + n_j + s_{ij})$ is the total population in these locations. We have $m_i = \overline{m}(1+\delta_i)$ and $n_j = \overline{m}(1+\delta_j)$, with $\delta_i = \dfrac{m_i - \overline{m}}{\overline{m}}$ and $\delta_j = \dfrac{n_j - \overline{m}}{\overline{m}}$. Equation (S7) then becomes ($\rho^2 = 1/C$)

$$E = \frac{1}{N} \sum_{\{i,j: i \neq j\}} \left[ \ln\left( \frac{\overline{m}^{\alpha+\beta}(1+\delta_i)^\alpha (1+\delta_j)^\beta}{\rho^2 r_{ij}^\gamma} \right) - \ln\left( \frac{\overline{m}^3 (1+\delta_i)^2 (1+\delta_j)}{\overline{m}^2 N_{ij}(N_{ij}-1-\delta_j)} \right) \right]^2.$$

$E$ reaches its minimum, resulting in a better fit, if each element in the sum is close to zero:

$$(\alpha+\beta-1)\ln(\overline{m}) + (\alpha-2)\ln(1+\delta_i) + (\beta-1)\ln(1+\delta_j) - \ln(\rho^2 r_{ij}^\gamma) + \ln(N_{ij}) + \ln(N_{ij}-1-\delta_j) = 0$$
for all $\{i,j : i \neq j\}$.

This is true if the following conditions are satisfied:

$(\alpha+\beta-1)\ln(\overline{m}) = 0 \qquad \Rightarrow \qquad \alpha+\beta = 1$ \hfill (i)

$(\alpha-2)\ln(1+\delta_i) + (\beta-1)\ln(1+\delta_j) = 0 \qquad \Rightarrow \qquad (\alpha-2)\delta_i + (\beta-1)\delta_j \approx 0$ \hfill (ii)

$\ln(N_{ij}) + \ln(N_{ij}-1-\delta_j) - \ln(\rho^2 r_{ij}^\gamma) = 0 \Rightarrow \ln(N_{ij}/\rho r_{ij}^{\gamma/2}) - \ln[(N_{ij}-1-\delta_j)/\rho r_{ij}^{\gamma/2}] = 0 \Rightarrow \rho r_{ij}^{\gamma/2} \approx N_{ij}$
\hfill (iii)

Equation (i) reproduces the result we obtain if we derive the gravity law from the radiation model in the limit of constant population density. Equation (ii) is always true if $\alpha = 2$ and $\beta = 1$, but this is in contrast with condition $\alpha+\beta = 1$ obtained from (i) and corresponding to the dominant contribution to $E$. Indeed, if the deviations of populations from the *local* average are small, i.e. $\delta_i, \delta_j \ll 1$, then the contribution to $E$ of (ii) is small and negligible compared to $\ln(\overline{m})$, which is the largest term if $\overline{m} \gg 1$. Condition (iii) is satisfied for $\gamma = 4$ when the locations are *regularly spaced* or have equal area, like in a grid. In this case $\rho$ can be interpreted as the global density of locations (i.e. the total number of locations divided by the country's area). If, on the contrary, the locations are not regularly spaced, the density of locations within origin *i* and destination *j* will be different for every pair (*i*, *j*), and thus it will be impossible to find a value for parameter $\rho$ that is constant throughout the country, compromising the goodness of the fit.

Thus, when locations have approximately the same area and the local deviations of population are small, Equations (i)-(iii) give the following theoretical predictions for the best-fit values: $\alpha + \beta = 1$ and $\gamma = 4$.

We now turn to the empirical data and show that the better Eqs. (i)-(iii) are satisfied, the better is the gravity law's fit. We will analyse the commuting fluxes within the United States, for which we have empirical data at the county level (i.e. the number of commuters between the U.S. counties), by grouping the counties into larger regions, and aggregating the fluxes between them accordingly. In particular, we will show two different aggregations, one that fulfils equations (i)-(iii) and one that does not, and we will compare the performance of the gravity law in the two cases.

*Locations with equal population*. The locations are the congressional districts, obtained by an appropriate aggregation of counties such that all districts have roughly the same population (see **Fig. S2**). In particular, there are 435 districts with an average population of $652,701 \pm 153,834$ individuals. This subdivision does *not* fulfil equation (iii), requiring regularly spaced locations. Indeed, as can be seen in the map of **Fig. S2**, the density of locations is higher in highly populated regions, where districts' average area is also very small, compared to districts in low populated regions.

The gravity law's best-fit parameters for commuting trips at the district level are $[\alpha, \beta, \gamma]$ = [-0.31, 0.31, 2.80] for $r < 500$ km, $[\alpha, \beta, \gamma]$ = [-0.23, -0.33, 0.14] for $r > 500$ km, that do not fulfill the conditions $\alpha + \beta = 1$ and $\gamma = 4$ obtained by equations (i)-(iii). It is interesting to note that $\alpha$ and $\beta$ are sometimes negative, meaning that when the population increases, less travelers are predicted.

To assess the goodness of the fit we calculate the error as the mean square deviation of the logarithms of trips, $E = \frac{1}{N} \sum_{\{i,j: i \neq j\}} [\ln(T_{ij}^{GM}) - \ln(T_{ij}^{data})]^2$, where the sum extends over all locations pairs $(i, j)$ such that $T_{ij}^{data}, T_{ij}^{GM} > 0$, and $N$ is the total number of county pairs considered. In this case the error $E^{district}$ is 1.43.

*Locations with equal area.* The locations are defined by placing on the U.S. map a square grid, and grouping together all counties whose geometrical centroids lie within the same square. In this way we obtain 400 equally spaced locations having roughly the same area, as can be seen in **Fig. S3**; the average area is $19,160 \pm 8,658$ km$^2$. This subdivision thus fulfils the requirement of equally spaced locations, necessary for Eq. (iii) to hold. The gravity law's best-fit parameters for commuting trips are $[\alpha, \beta, \gamma] = [0.48, 0.58, 4.35]$ for $r < 390$ km, $[\alpha, \beta, \gamma] = [0.40, 0.33, 0.49]$ for $r > 390$, that are compatible with the conditions $\alpha + \beta = 1$ and $\gamma = 4$ given by equations (i)-(iii) when $r < 390$ km (where 95% of trips takes place). When $r > 390$ km the best fit parameters do not agree with our estimates, due to the difficulty to predict the small fraction of long-distance trips, as discussed in the main text. The error is $E^{grid} = 0.96$, smaller than in the previous case, confirming that the gravity law works better when locations have equal area and are equally spaced, and as long as there is a local uniformity in the population distribution. This result is also an indirect validation of the radiation model, because in order to obtain the conditions $\alpha + \beta = 1$ and $\gamma = 4$ that agree with the experimental values, we assumed $T_{ij}^{data} \approx T_{ij}^{Rad}$.

## 7   Differences between the Radiation model and other decision-based models.

The Radiation model's approach is similar to the original idea behind the intervening-opportunities model[26,38], but the developments and the final results and performances are radically different.

The intervening-opportunities (IO) model proposes that "the number of persons going a given distance is directly proportional to the number of opportunities at that distance and inversely proportional to the number of intervening opportunities"[26]. Assuming that the number of opportunities in a location is proportional to its population, we have that the number of trips between locations $i$ and $j$ is $T_{ij} \propto n_j / (s_{ij} + m_i)$ where $n_j$ is population in location $j$ and ($s_{ij} + m_i$) is the population in all locations between $i$ and $j$. In **Fig. S5** we show the performance of Stouffer's IO model tested on US commuting fluxes.

This first formulation was subsequently recast in a stochastic approach that has become the standard theory of IO models, and it is defined as follows[38]: the probability that a trip will terminate in location $j$ is equal to the probability that $j$ contains an acceptable destination and that an acceptable destination closer to the origin $i$ has not been found. The number of trips is shown to be $T_{ij} \propto \left[ e^{-\lambda(s_{ij}+m_i)^\alpha} - e^{-\lambda(s_{ij}+m_i+n_j)^\alpha} \right]$, where $e^{-\lambda}$ is the probability that a single opportunity is not sufficiently attractive as destination, and $\lambda$ and $\alpha$ are fitting parameters.

The IO model shares with the radiation model the intuition of using distance to sort the possible destinations, identifying that the correct variable to calculate $T_{ij}$ is $s_{ij}$ and not $r_{ij}$ as used in the gravity law. Besides this, the IO model suffers from problems that are similar to those highlighted earlier for the gravity law. In particular, the lack of universal values for parameters $\lambda$ and $\alpha$, and the difficulty to calibrate these values makes impossible to obtain the correct number of total trips[28,39,40]. These shortcomings, along with a higher computational complexity, have led to the success of the gravity law over the IO model in recent years[40].

Another decision-based approach is the random utility (RU) model[27,41]. In RU models, utilities distributed according to a logit-function are assigned to each destination. Utilities can depend on various variables including travel times, distances, job opportunities. A rationality paradigm is used to model the decision process: the individual will choose the alternative that maximizes a random utility function subject to specific cost constraints. Although RU model's predictions can be derived from first principles, they depend on tunable parameters that are context specific, and do not clearly outperform the gravity law's estimates[40].

## 8   Self-similarity in human mobility

The radiation model helps us uncover a previously unsuspected scale-invariance in commuting patterns. Let us denote with $p_s(>s|m)$ the probability of a trip from a location with population $m$ to any destination that lies beyond a circle of radius $r(s)$ centered at the source (i.e. a circle containing population $s$). The radiation model (2) predicts that this probability has the simple form

$$p_s(>s|m) = \frac{1}{1+s/m}, \qquad (S9)$$

which is scale invariant under the transformation $m \to \lambda m$ and $s \to \lambda s$. Such scale-invariance is absent, however form the gravity law (1). To offer empirical evidence for this statistical self-similarity in **Fig. 4a** we plot the $s$-dependence of $p_s(>s|m)$ for 2,000

US counties, the obtained curves spanning about six orders of magnitude in *s*. Yet, if we plot $p_s(>s|m)$ in function of the dimensionless *s/m*, as predicted by (S9), the differences between the different curves decrease significantly, and the universal function 1/(1+*x*) predicted by (S9) offers a close lower bound to the rescaled curves (**Fig. 4b**).

### 9 Beyond the radiation model

Here we show how Eq. (2) is not a rigid end point of our approach, but can serve as a platform that can be improved upon in specific environments. For example, the radiation model slightly underestimates the number of very long commuting patterns (those over 1000 km, **Fig. 2a** and **Fig. S7**). While this impacts less than 1% of all trips (336,113 of 34,116,820 travelers), the agreement could be improved via an effective distance, used in transportation modeling[42], defining the variable $s_{ij}$ as the population within an *effective* travel time from the origin. Indeed, if efficient means of transport are available (like direct flights), individuals are willing to commute farther.

As a second example we consider the fact that an individual has a home-field advantage when searching for jobs in his/her home county, being more familiar with local employment opportunities. This is justified considering that individuals have a better knowledge of their home location, because they know more people, have more connections, and thus have access to more employment opportunities than in an unfamiliar location, where they can typically apply only to widely advertised jobs.

In the radiation model we implement this home-advantage in the following way: while for the other locations the number of job offers is proportional to the respective populations, for the home location the number of job opportunities is proportional to the population, $m/n_{jobs} + \varepsilon/n_{jobs}$, which corresponds to the effective the addition of $\varepsilon$ people. Below we show the new expressions of the probability of one trip, $P(1|m,n,s)$, and the probability of a trip to a destination beyond a circle of radius $r(s)$ centered at the origin, $p_s(>s|m)$, for the home-advantage variant described above:

$$P(1|m,\varepsilon,n,s) = \frac{(m+\varepsilon)n}{((m+\varepsilon)+s)((m+\varepsilon)+n+s)} \quad \text{(S10 a)}$$

$$p_s(>s|m,\varepsilon) = \frac{(m+\varepsilon)}{(m+\varepsilon)+s} \quad \text{(S10 b)}$$

The adjusted law is now invariant under the $(m+\varepsilon) \to \lambda(m+\varepsilon)$ and $s \to \lambda s$ transformation. This version of the radiation model, with $\varepsilon = 35,000$, provides a better collapse of the rescaled distributions $p_s(>s|m)$, as shown in **Fig. 4c**. In **Fig. S6** we show that the new version also provides comparable results on the prediction of $p(T|m,n,r)$, $P_{dist}(r)$, and $P_{dest}(n)$ distributions, for which the original radiation model was already over-performing the gravity model. Moreover, the better performance of the new version is particularly evident in the pairwise comparison of real data fluxes and the radiation model's predictions, shown in the scatter plot of **Fig. S6c** (the individual points are omitted for the sake of clarity). The error bars corresponding to the new version are systematically shorter than the error bars of the original radiation model, indicating smaller fluctuations of its predictions and demonstrating that Eq. (2) is not a rigid end point of our approach, but can serve as a platform that can be improved upon in specific environments.

**Additional References**

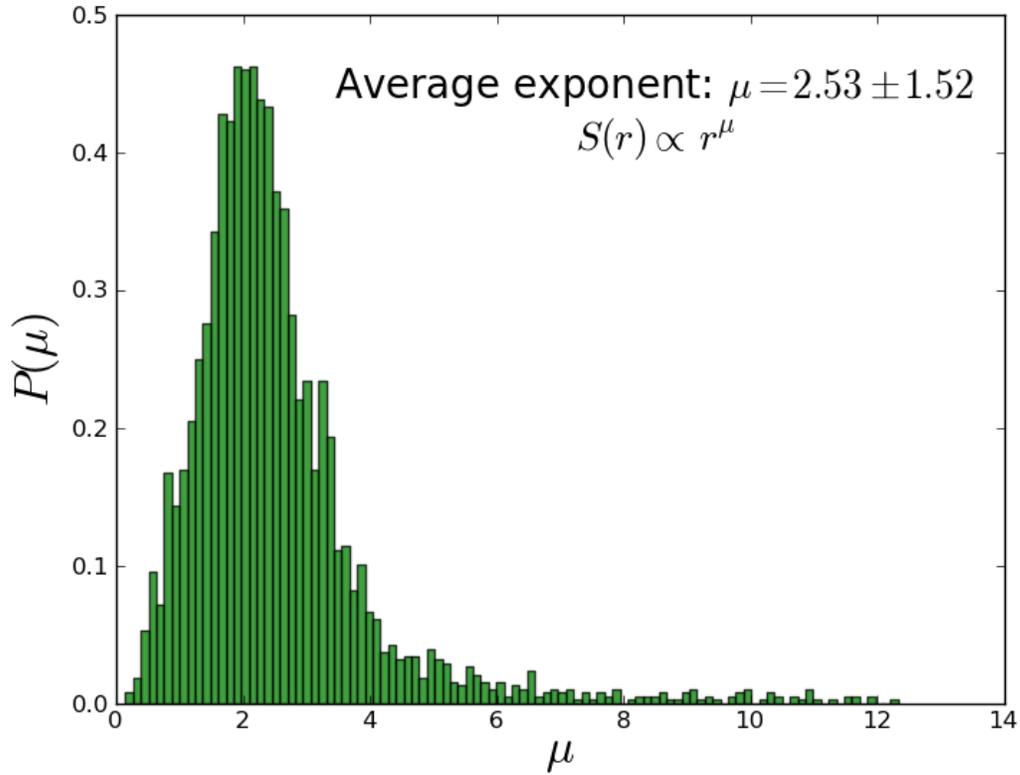

**Figure S1**

Figure S1. The distributions of exponents $\mu$, obtained by fitting the data with $s(r) \propto r^{\mu}$, starting from each US county. $P(\mu)$ measures the degree of homogeneity of the population distribution in the region of short trips, $r < 119$ m, capturing about 95% of all trips. We fitted the $r$ vs $s$ scaling relationship within the mentioned cutoff separately for each county, measuring the total population, $s$, within a circle of radius $r$ centered in the location, as $r$ increases. We then calculated $P(\mu)$, the distribution of the fitted exponents $\mu$ of all locations. We found a high heterogeneity, the mean value (with a large variance) being $\mu = 2.53$ and not 2, as we would expect for a uniform spatial population distribution.

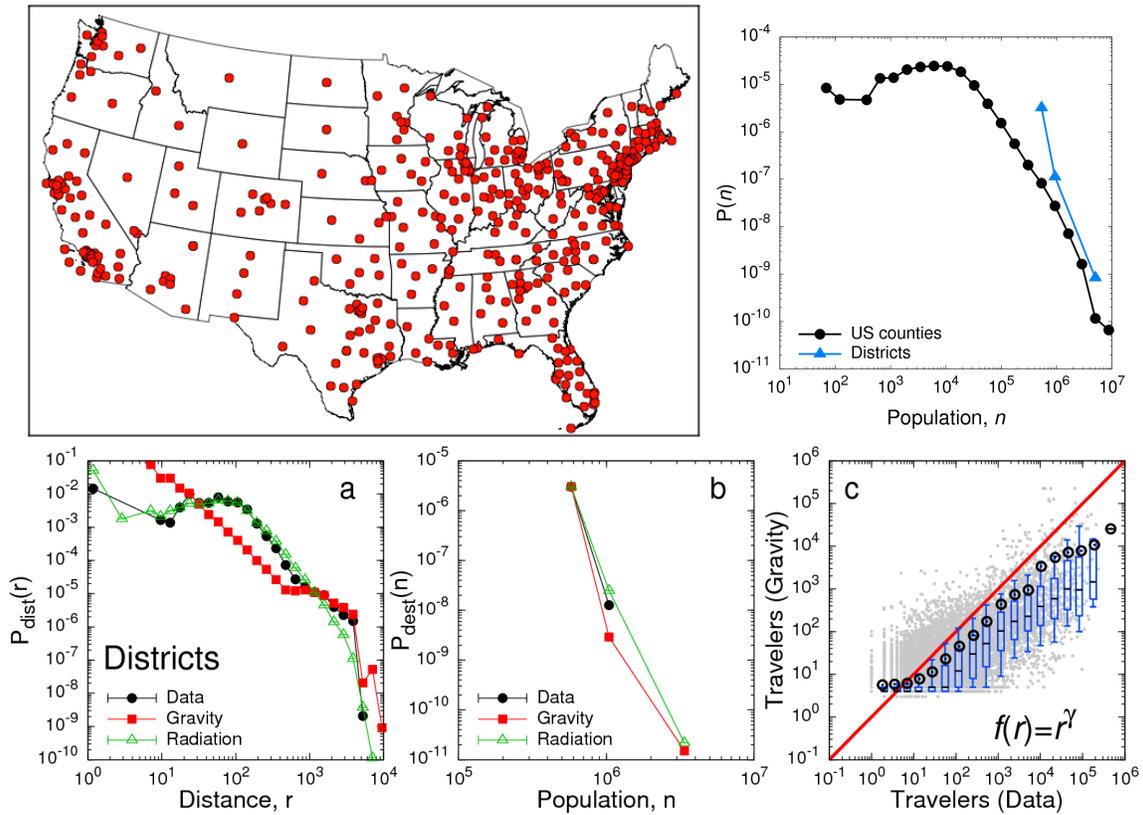

**Figure S2**

Figure S2. Commuting trips among 435 congressional districts, which are regions of *equal population* (the population distribution $P(n)$, top right corner, is highly peaked around the average 652,701±153,834). Best-fit parameters: $[\alpha, \beta, \gamma]$ = [-0.31, 0.31, 2.80] for $r < 500$ km, $[\alpha, \beta, \gamma]$ = [-0.23, -0.33, 0.14] for $r > 500$ km. The scatter plot of panel *c* shows that the predicted trips do not agree with the experimental data because they do not align over the red line *y=x*. The poor quality of the fit also is noticeable in panel *a* where the distribution of trip lengths $P_{dist}(r)$ is displayed: the red squares, corresponding to the gravity fit, disagree with the distribution from the data. The fit error is $E^{district} = 1.43$.

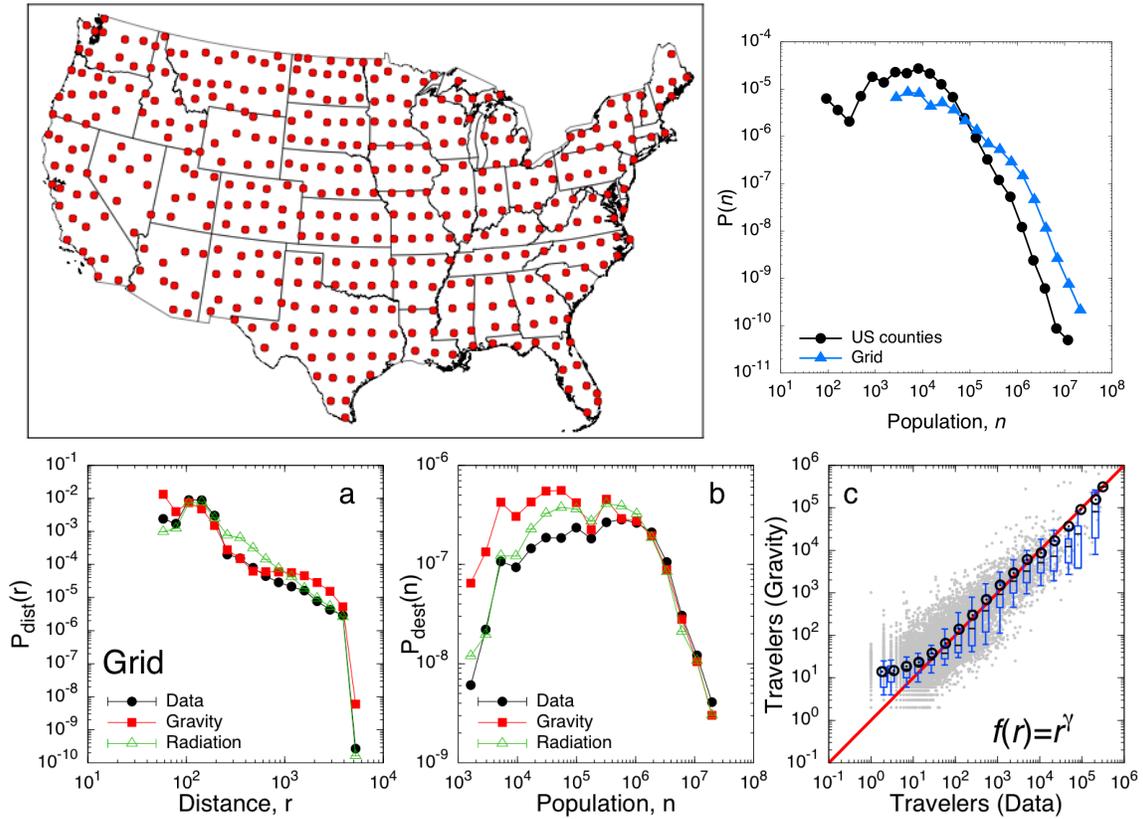

**Figure S3**

Figure S3. Commuting trips among 400 regions of *equal area* displaced as a grid. Best-fit parameters: [$\alpha, \beta, \gamma$] = [0.48, 0.58, 4.35] for $r < 390$ km, [$\alpha, \beta, \gamma$] = [0.40, 0.33, 0.49] for $r > 390$ km. In this case the scatter plot in panel *c* shows a much better agreement between data and the gravity prediction: the points are closely aligned along the red line *y=x*. This is confirmed in panels *a* and *b*, where the distributions of trips lengths and population in the destinations of the gravity law (red squares) are close to the distributions of data (black circles). The fit error is, accordingly, less than in the previous case $E^{grid} = 0.96$.

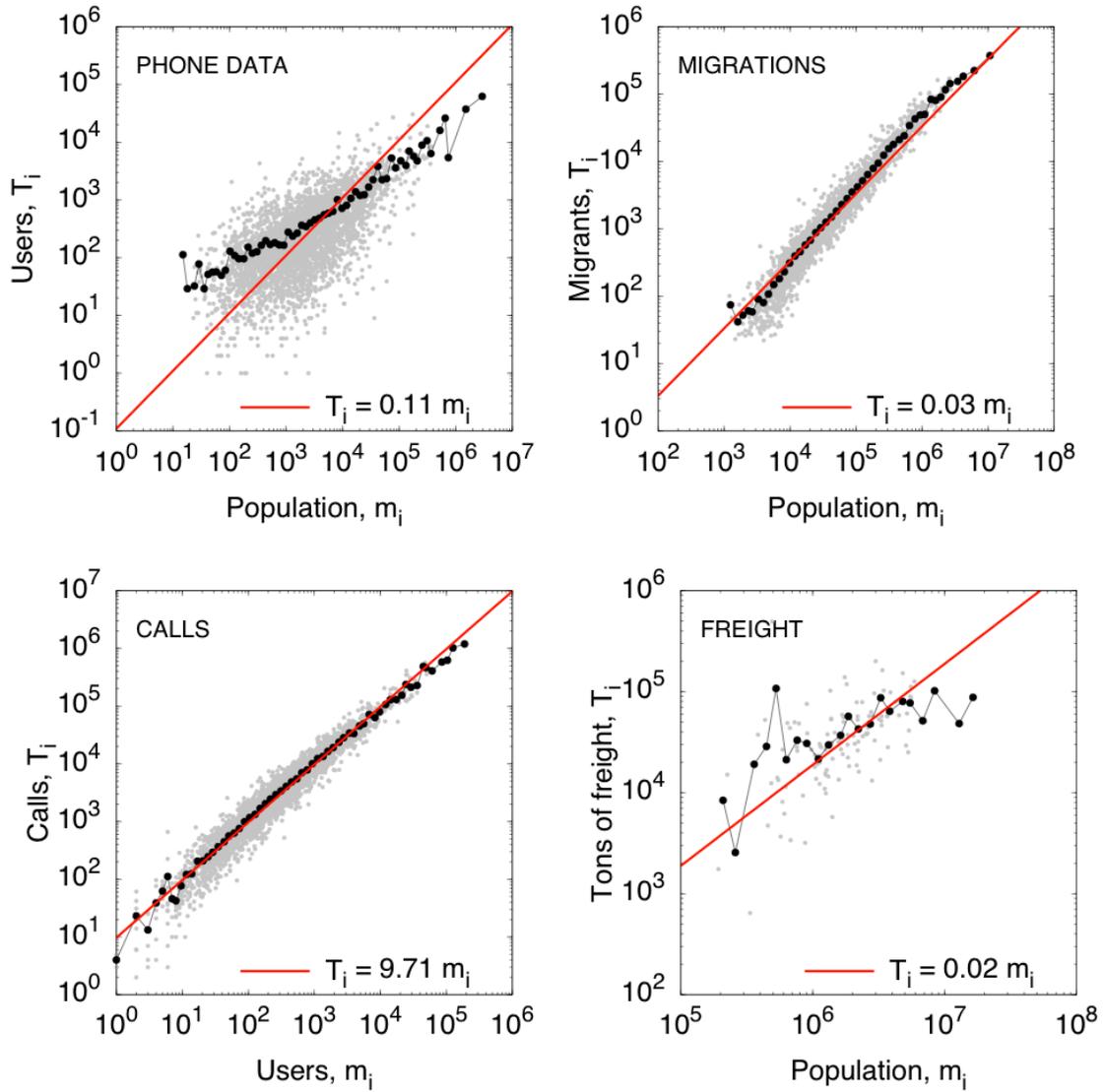

**Figure S4**

Figure S4. The number of emitted particles, $T_i \equiv \sum_{j \neq i} T_{ij}$, versus the population of each location, $m_i$. The assumption $T_i \propto m_i$ is verified for all cases except for freight transportation and phone data, because the weight of commodities produced in a region is not related to the region's population, and the cell phone company's market share is not uniformly distributed in the country. For these two cases we used the measured $T_i$.

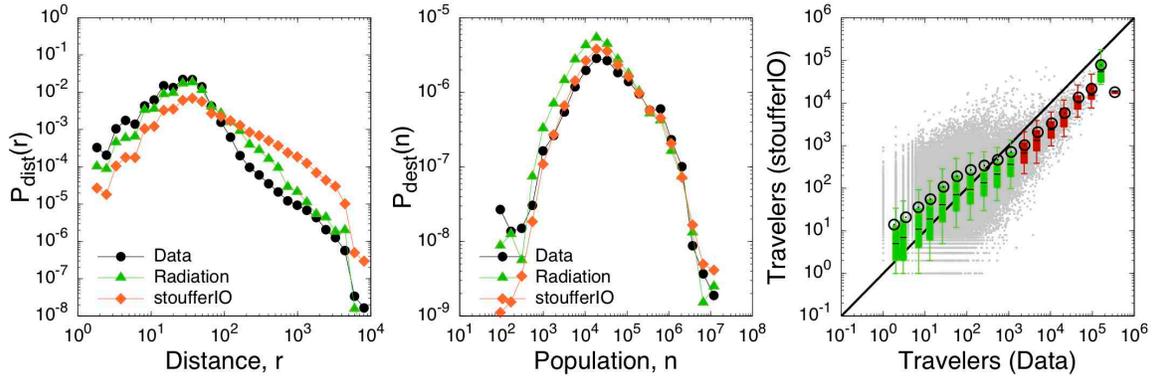

**Figure S5**

Figure S5. Comparison between the Radiation model and Stouffer's original Intervening Opportunities model. Here we display the performance of Stouffer's IO model[26] applied to the US commuting trips (the same kind of plots of Fig. 2e,f,d). While the IO model provides a good estimate of $P_{dest}(n)$ (center plot), it is unable to reproduce the $P_{dist}(r)$ distribution at large distances (left plot), and systematically underestimates the high fluxes observed in the data (right plot).

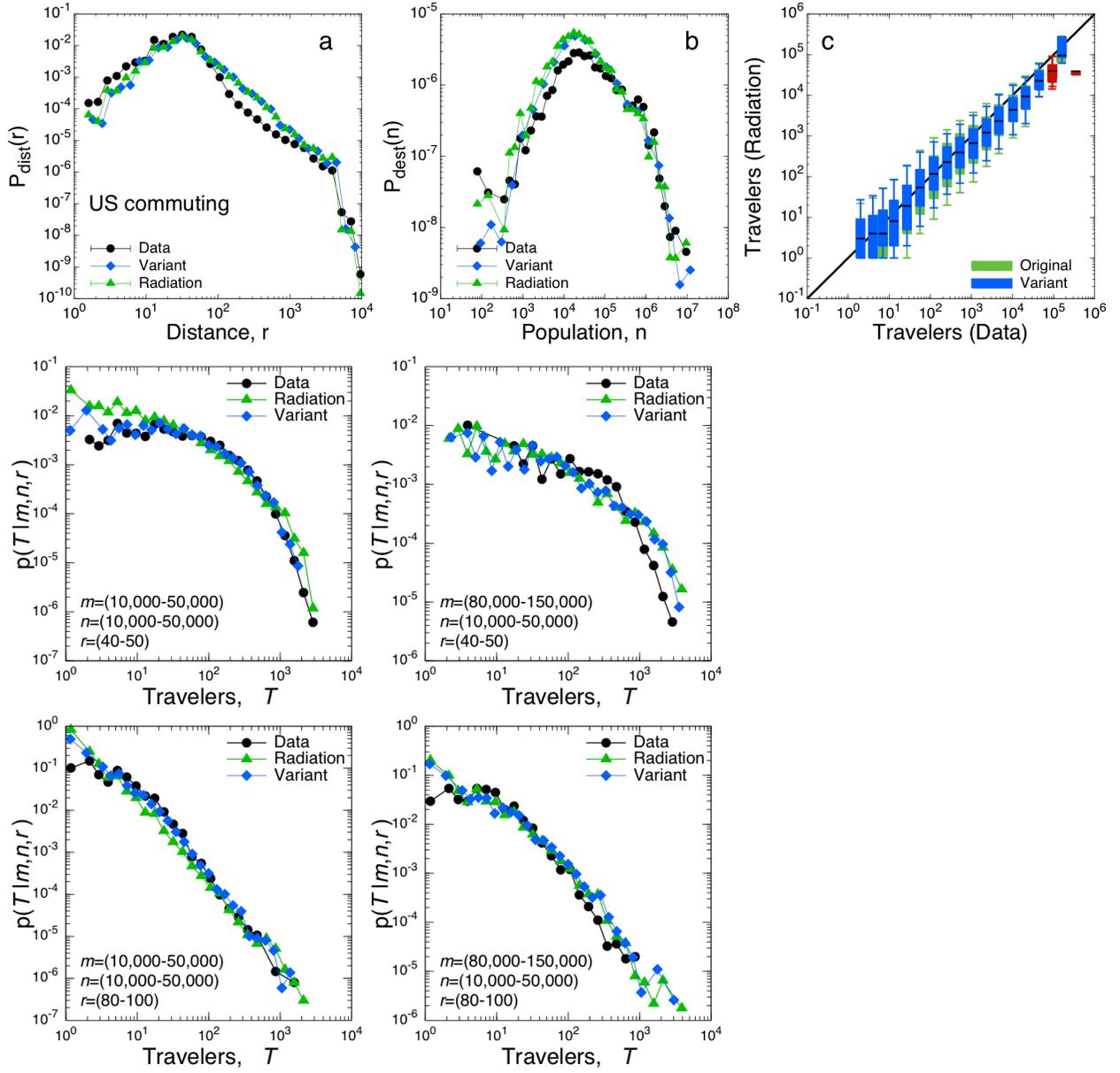

**Figure S6**

Figure S6. Variant of the radiation model. See section 9, and Fig. 2.

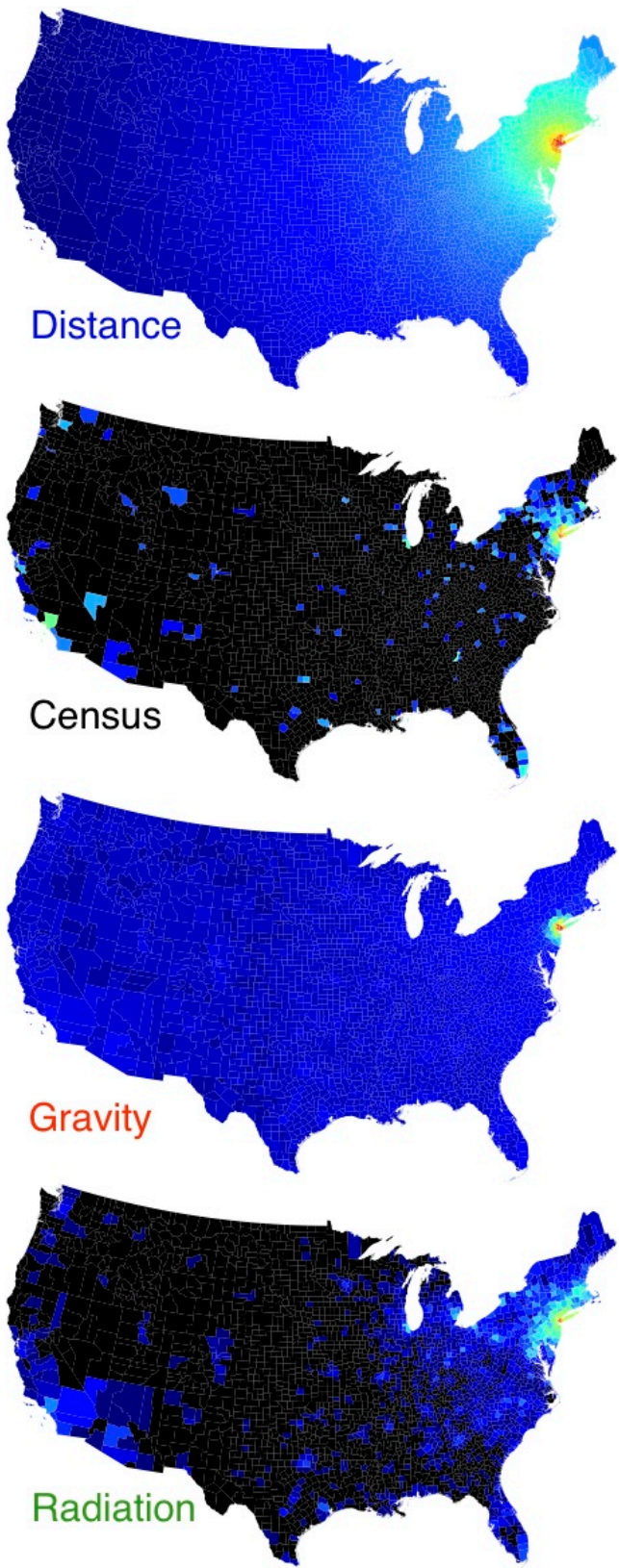

Figure S7. Trips from New York County (FIPS 36061). In these four figures we colored the US counties according to the distance (same as Fig. 4a,d in the main text) (top), and number of commuters arriving from New York County, as estimated by the US Census 2000 (second from top), the gravity law with fitting parameters of Ref. 14 (center), and the radiation model (third panel). The color scale is logarithmic: red is the highest flux (273,244 travelers) (and shortest distance), and dark blue the lowest flux (1 traveler) (and highest distance); while black corresponds to an estimate of less than one traveler. We note that while the gravity law overestimates most fluxes, the predictions of the radiation model are rather similar to the data, with two important exceptions: i) it underestimates distant trips to large cities (like Seattle, Las Vegas, Miami), and ii) it does not detect the effect of "state boundaries": indeed, real data indicate that new yorkers are more likely to work inside their state (New York State) (73%) than what predicted by the radiation model (58%).

**Figure S7**

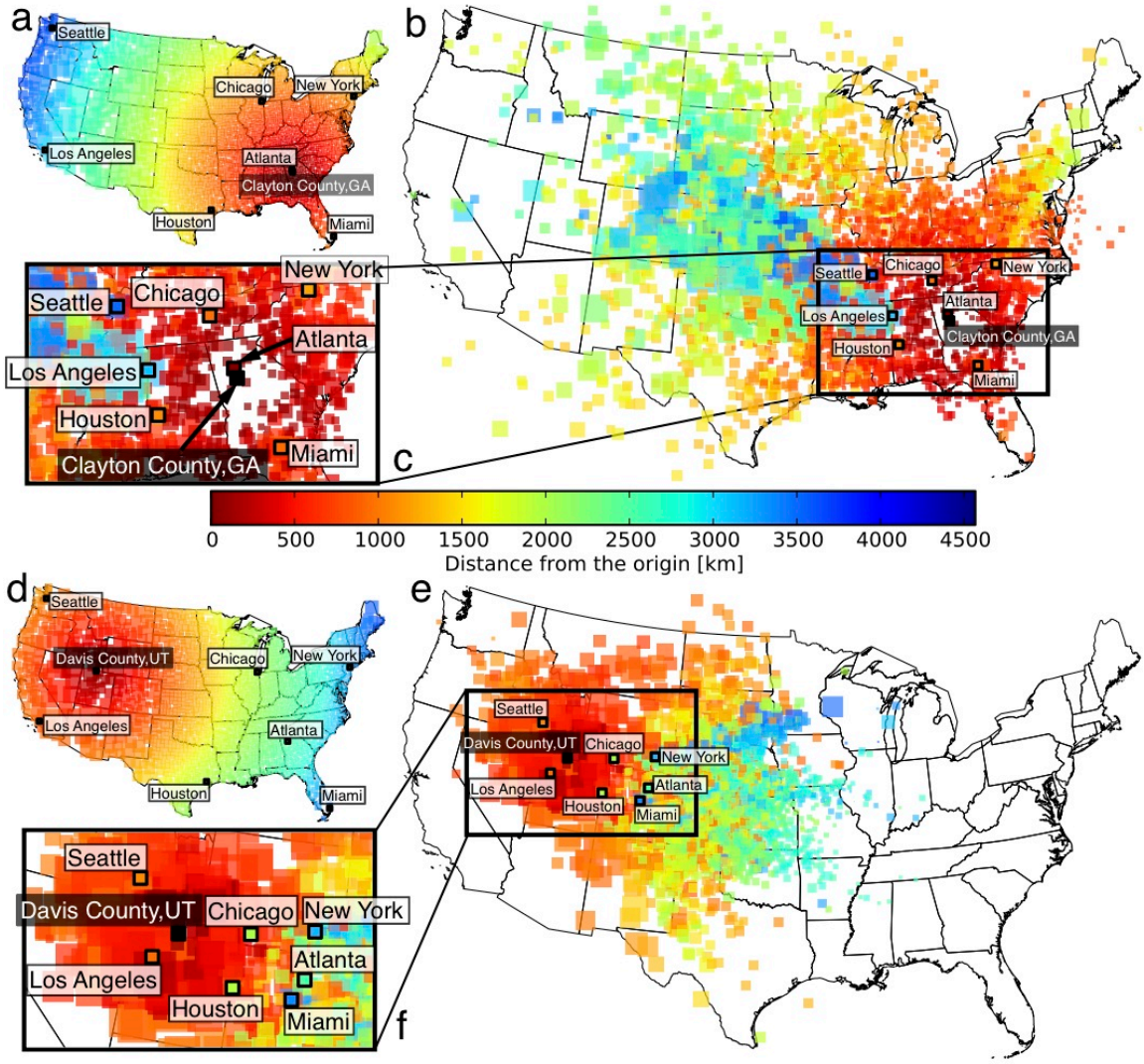

**Figure S8**

Figure S8. **Commuting landscapes.** The commuting-based attractiveness of various US counties as seen from the perspective of an individual in (*a-c*) Clayton County, GA, and (*d-f*) Davis County, UT. (*a,d*) Distance of all US counties from Clayton county, GA, in (*a*), and Davis county, UT, in (*d*). Each square represents a county, whose color denotes the distance from Clayton (Davis) county and the size is proportional to the county's area. Seven large cities are shown to guide the eye. *b,c* (*e,f*), The distance of counties relative to Clayton, GA (Davis, UT) has been altered to reflect the likelihood that an individual from Clayton (Davis) county would commute to these, as predicted by (2). Big cities appear much closer than suggested by their geographic distance, due to the many employment opportunities they offer. For the UT-based individual the US seems to be a "smaller" country, than for the GA-based individual. Indeed, given the low population density surrounding Davis county, the employee must travel far to satisfy his/her employment needs. Equally interesting is the fact that the Clayton county, GA, based individual sees an effective employment "hole" in its vicinity (*c*). The reason is the nearby Atlanta, which offers so many employment opportunities, that all other counties become far less desirable.
45